\documentclass[11pt]{article}

\usepackage{amsmath}
\usepackage{amssymb}
\usepackage{extarrows}
\usepackage{graphicx}
 \usepackage{indentfirst}
\usepackage{hyperref}

\renewcommand{\baselinestretch}{1.1}
  \renewcommand{\arraystretch}{1.0}
\begin{document}

 \title{Comment  on Two schemes for Secure\\
Outsourcing of Linear Programming}

 \author{Zhengjun Cao$^1$ and  Lihua Liu$^2$}

     \footnotetext{$^1$Department of Mathematics, Shanghai University, Shanghai,
  China.  \\
     $^2$Department of Mathematics, Shanghai Maritime University,
  China.
   \textsf{liulh@shmtu.edu.cn}
    }

 \date{}\maketitle

\begin{quotation}
 \textbf{Abstract}. Recently, Wang et al. [IEEE INFOCOM 2011, 820-828], and Nie et al. [IEEE AINA 2014, 591-596] have proposed two schemes for secure
outsourcing of large-scale linear programming (LP).  They did not consider the standard form:
minimize $\textbf{c}^{T}\textbf{x},$ subject to $\textbf{A}\textbf{x} = \textbf{b}, \textbf{x}\geq 0$.
 Instead, they studied  a peculiar form: minimize $\textbf{c}^{T}\textbf{x},$ subject to $\textbf{A}\textbf{x} = \textbf{b}, \textbf{B}\textbf{x}\geq 0$,  where $\textbf{B}$ is a non-singular matrix.
 In this note, we stress  that the proposed peculiar form is unsolvable and meaningless.  The two schemes  have confused the \emph{functional inequality constraints} $\textbf{B}\textbf{x}\geq 0 $ with the \emph{nonnegativity constraints} $\textbf{x}\geq 0 $ in the linear programming model. But the condition $\textbf{x}\geq 0 $ is indispensable to the simplex method. Therefore, both two schemes failed.

 \end{quotation}

\section{Introduction}

Recently,  Wang et al. \cite{WRW11}, and Nie et al. \cite{N14} have proposed two schemes for secure
outsourcing of large-scale linear programming (LP).  They did not consider the standard form:
$$ \mbox{minimize}\quad \textbf{c}^{T}\textbf{x}, \quad \mbox{subject to}\quad \textbf{A}\textbf{x} = \textbf{b}, \textbf{x}\geq 0,$$
 Instead, they studied  a peculiar form:
 $$\mbox{minimize}\quad \textbf{c}^{T}\textbf{x}, \quad \mbox{subject to} \quad \textbf{A}\textbf{x} = \textbf{b}, \textbf{B}\textbf{x}\geq 0$$
  where $\textbf{A}$ is an $m\times n$ matrix, $\textbf{c}$ is an $n\times 1$ vector,   $\textbf{b}$ is an $m\times 1$ vector,  $\textbf{x}$ is an $n\times 1$ vector of variables, and $\textbf{B}$ is an $n\times n$ non-singular matrix.

 In this note, we would like to stress  that the proposed peculiar form is unsolvable and meaningless.  The two schemes  have confused the \emph{functional inequality constraints} $\textbf{B}\textbf{x}\geq 0 $ with \emph{nonnegativity constraints} $\textbf{x}\geq 0 $ in the linear programming model. In nature,  the condition $\textbf{x}\geq 0 $ is indispensable to the simplex method. Thus, both two schemes failed.

\section{Preliminaries}

The \emph{standard form} for a linear programming problem can be described as follows. Select the values for $x_1, \cdots, x_n$ so as to
$$\mbox{maximize}\quad c_1x_1+c_2x_2+\cdots+c_nx_n, $$
subject to the restrictions
\begin{eqnarray*}
 a_{11}x_1+a_{12}x_2+\cdots+a_{1n}x_n&\leq& b_1  \\
 a_{21}x_1+a_{22}x_2+\cdots+a_{2n}x_n&\leq& b_2  \\
&\vdots& \\
 a_{m1}x_1+a_{m2}x_2+\cdots+a_{mn}x_n&\leq& b_m
\end{eqnarray*}
and
$$x_1\geq 0, x_2\geq 0, \cdots, x_n\geq 0. $$
$c_1x_1+c_2x_2+\cdots+c_nx_n$ is called the \emph{objective function}. The first $m$ constraints are sometimes called \emph{functional constraints}. The $x_j\geq 0$ restrictions are called \emph{nonnegativity constraints}.

The simplex method, a general procedure for solving linear programming problems, is based on solving systems of equations. Therefore, it has to firstly convert the functional inequality constraints to \emph{equivalent equality constraints}. This conversion is accomplished by introducing \emph{slack variables}. After the conversion, the original linear programming model can now be replaced by the equivalent model (called the \emph{augmented form}).

Using matrices, the standard form for the general linear programming model becomes
$$\mbox{maximize}\quad \textbf{c}^{T}\textbf{x}, \quad \mbox{subject to} \quad \textbf{A}\textbf{x} \leq \textbf{b},  \textbf{x}\geq 0$$
  where $\textbf{A}$ is an $m\times n$ matrix, $\textbf{c}$ is an $n\times 1$ vector,   $\textbf{b}$ is an $m\times 1$ vector,  and $\textbf{x}$ is an $n\times 1$ vector of variables. To obtain the augmented form of the problem, introduce the column vector of slack variables $\textbf{x}_s=(x_{n+1}, \cdots, x_{n+m})^{T}$ so that the constraints become
  $$[\textbf{A}, \textbf{I}]  \left[
              \begin{array}{c}
                \textbf{x} \\
                \textbf{x}_s \\
              \end{array}
            \right]=\textbf{b} \quad\mbox{and}\ \left[
              \begin{array}{c}
                \textbf{x} \\
                \textbf{x}_s \\
              \end{array}
            \right]\geq \textbf{0},  $$
   where $\textbf{I}$ is the $m\times m$ identity matrix, and the null vector $\textbf{0}$ now has $n+m$ elements.

Notice that the nonnegativity constraints are left as inequalities because they are used to determine the \emph{leaving basic variable} according to the \emph{minimum ratio test} \cite{HL9}.

\section{Analysis of the two schemes for secure
outsourcing of LP}

\subsection{Review}

We now take the scheme in Ref.\cite{WRW11} as the example to show the incorrectness of the proposed peculiar form (see the page 822 of Ref.\cite{WRW11}, and the page 592 of Ref.\cite{N14}). In the scheme, there are two entities, the client and the server. The client has the original problem  $$\mbox{minimize}\quad \textbf{c}^{T}\textbf{x}, \quad \mbox{subject to} \quad \textbf{A}\textbf{x} = \textbf{b},\ \textbf{B}\textbf{x}\geq 0 \eqno(1)$$
  where $\textbf{A}$ is an $m\times n$ matrix, $\textbf{c}$ is an $n\times 1$ vector,   $\textbf{b}$ is an $m\times 1$ vector,  $\textbf{x}$ is an $n\times 1$ vector of variables, $\textbf{B}$ is an $n\times n$ non-singular matrix.

To ensure the privacy of input and output, the client transforms the original problem into the following problem
 $$\mbox{minimize}\quad \textbf{c}'^{T}\textbf{y}, \quad \mbox{subject to} \quad \textbf{A}'\textbf{y} = \textbf{b}', \ \textbf{B}'\textbf{y}\geq 0 \eqno(2)$$
  where
  $$
  \left\{
    \begin{array}{l}
      \textbf{A}'=\textbf{Q}\textbf{A}\textbf{M} \\
      \textbf{B}'=(\textbf{B}-\textbf{P} \textbf{Q}\textbf{A})\textbf{M} \\
      \textbf{b}'=\textbf{Q}(\textbf{b}+\textbf{A}\textbf{r}) \\
      \textbf{c}'=\gamma \textbf{M}^{T} \textbf{c} \\
      \textbf{y}=\textbf{M}^{-1}(\textbf{x}+\textbf{r}) \\
    \end{array}
  \right. $$
satisfying
$$
|\textbf{B}'| \neq 0,  \textbf{P}\textbf{b}'=\textbf{B}\textbf{r}, \textbf{b}+\textbf{A}\textbf{r}\neq 0, \gamma>0,
$$
where $\textbf{P}$ is an $n\times m$ matrix,  $\textbf{Q}$ is a random $m\times m$ non-singular matrix,  $\textbf{M}$ is a random $n\times n$ non-singular matrix,  and $\textbf{r}$ is an $n \times 1$ vector.

The client then sends the problem (2) to the server, instead of the original problem (1).
\subsection{Analysis}

When the server receives the problem (2),  he has to introduce  the nonnegativity conditions $\textbf{y}\geq 0$ into it and solve the following problem $$\mbox{minimize}\quad \textbf{c}'^{T}\textbf{y}, \quad \mbox{subject to} \quad \textbf{A}'\textbf{y} = \textbf{b}',\ \textbf{B}'\textbf{y}\geq 0, \ \textbf{y}\geq 0 \eqno(3)$$
This is because the constraints $\textbf{B}'\textbf{y}\geq 0$ should be viewed as a part of the functional constraints, not the necessary  nonnegativity constraints. Unless $(\textbf{B}-\textbf{P} \textbf{Q}\textbf{A})\textbf{M}$ can be rewritten as a diagonal matrix where the entries on the main diagonal are strictly positive (in such case, $\textbf{B}'\textbf{y}\geq 0$ implies $\textbf{y}\geq 0$). 

Unfortunately,  the solution of the following problem
$$\mbox{minimize}\quad \textbf{c}^{T}\textbf{x}, \quad \mbox{subject to} \quad \textbf{A}\textbf{x} = \textbf{b}, \ \textbf{B}\textbf{x}\geq 0, \ \textbf{x}\geq 0 \eqno(4) $$
 cannot be derived from the solution of the problem (3), because  the transformation
$$\textbf{y}=\textbf{M}^{-1}(\textbf{x}+\textbf{r}), \quad \mbox{where} \quad \textbf{x}\geq 0 $$
cannot ensure that $\textbf{y}\geq 0$.

\emph{Remark 1}.  The authors of \cite{WRW11,N14} have confused the functional inequality constraints $\textbf{B}\textbf{x}\geq 0$  with the nonnegativity constraints $\textbf{x}\geq 0$. In fact, the proposed form is meaningless and unsolvable, unless $\textbf{B}\textbf{x}\geq 0$ can be rewritten as $\textbf{x}\geq 0$.

\section{Conclusion}
  We would like to restate that the procedure for determining the leaving basic variable in the simplex method requires that all the variables have nonnegativity constraints. One must draw a clear distinction between the functional inequality constraints and the nonnegativity constraints. 
  Notice that deriving  the augmented form of a standard form for a linear programming problem is very easy. It can be solely done by the client himself  even though  who is assumed to be of weak computational capability.

\end{document}